\documentclass[journal=jacsat,manuscript=article]{achemso}

\usepackage[version=3]{mhchem} 
\usepackage[utf8]{inputenc}
\usepackage[T1]{fontenc}
\usepackage{lineno}
\usepackage{amsmath}
\usepackage{subcaption}
\usepackage{graphicx}
\usepackage{mhchem}
\usepackage{siunitx}
\usepackage{hyperref}
\usepackage{chemformula} 
\usepackage{braket}
\usepackage{booktabs}

\author{Mathias Hilfiker}
\affiliation[Unknown University]
{Department of Physics and Materials Science, University of Luxembourg, L-1511 Luxembourg City, Luxembourg.}
\alsoaffiliation[BigPharma]
{Molecular AI, Discovery Sciences, R\&D, AstraZeneca, Gothenburg, Sweden}

\author{Leonardo Medrano Sandonas}
\affiliation{Institute for Materials Science and Max Bergmann Center for Biomaterials, TUD Dresden University of Technology, 01062 Dresden, Germany.}
\author{Alexandre Tkatchenko}
\affiliation[Unknown University]
{Department of Physics and Materials Science, University of Luxembourg, L-1511 Luxembourg City, Luxembourg.}

\author{Ola Engkvist}
\affiliation[BigPharma]
{Molecular AI, Discovery Sciences, R\&D, AstraZeneca, Gothenburg, Sweden}
\alsoaffiliation[chalmers]
{Department of Computer Science and Engineering, Chalmers University of Technology and University of Gothenburg, Gothenburg, Sweden}
\author{Marco Klähn}
\email{marco.klahn@astrazeneca.com}
\affiliation[BigPharma]
{Molecular AI, Discovery Sciences, R\&D, AstraZeneca, Gothenburg, Sweden}

\title[An \textsf{achemso} demo]
  {Machine-Learned Electrostatic Potentials for Accurate Hydration Free Energy Calculations}

\abbreviations{IR,NMR,UV}
\keywords{American Chemical Society, \LaTeX}

\begin{document}

\begin{abstract}
  Free energy calculations are widely used tools in computational chemistry, but their dependence on the assignment of partial charges during force field parametrization reduces their accuracy and reproducibility. In this work, we highlight the direct connection between the low accuracy of AM1-BCC charges on polar species and the poor accuracy of corresponding hydration free energy calculations. We then propose an XGBoost regressor trained on atomic descriptors to rapidly predict charges obtained with high-fidelity density functional theory calculations at PBE0-D3(BJ)/def2-TZVP level. The more accurate electrostatic description results in more reliable free energy calculations than those obtained with semi-empirical AM1-BCC charges. Finally, we leverage this predictive model in combination with a 1 ns gas-phase molecular dynamics simulation to propose the Boltzmann Percentile method for assigning charges representative of the conformational ensemble of a molecule. Charges obtained with this method are robust to different input conformations, and the resulting free energies, calculated on a subset of the FreeSolv dataset, show a root mean squared error of $1.69\ \text{kcal/mol}$ against the $3.05\ \text{kcal/mol}$ obtained with semi-empirical charges as well as a significantly better ranking.
Our method is easily integrable in the traditional workflow and requires the same computational resources. These two aspects make it a realistic tool for enhancing already expensive free energy calculations, and more in general, molecular dynamics simulations in condensed phase.
\end{abstract}

\section{Introduction}
Free energy (FE) calculations are computational physics-based methods that have found numerous applications in fields such as drug discovery\cite{cournia2024applications,shirts2012best,cournia2017relative,qian2024alchemical,shirts2010free}, where they are widely used in hit finding and lead optimization\cite{cournia2020rigorous}, and material science\cite{rickman2002free}. Their success is due to their ability to compute free energy differences between thermodynamic states, which ultimately allow scientists to estimate the feasibility of many natural processes such as solvation, protein-ligand binding, and polymerization. In any FE study, the preparation of the system and the analysis of the performed calculations require great attention, and while for the latter, well-established best practices exist\cite{mey2020best,klimovich2015guidelines}, it is still not completely clear the effect that the system's preparation has on downstream calculations. 

A key part of FE calculations are molecular dynamics (MD) simulations. Given the central role of electrostatic interactions in these simulations, typically described by Coulomb potentials, there has been considerable interest in how they are parametrized in classical force fields (FF). Coulomb potentials are often parameterized by using atomic partial charges. Previous studies have shown that solvation free energies are highly sensitive to the parametrization of the solute charges, both in the choice of the charge scheme\cite{jambeck2013partial, mobley2007comparison} and in the input conformation used to generate the charges\cite{reynolds1992errors, osato2025evaluating}. These works demonstrate that even small differences in assigned charges can lead to large variations in the computed energies, ultimately compromising the reproducibility and accuracy of the FE calculations. Recently, the MACE-OFF machine learning force field (MLFF)\cite{moore2024computing,batatia2022mace,kovacs2025mace} has been used to perform hydration free energy calculations of 36 compounds of the FreeSolv dataset\cite{mobley2014freesolv} with outstanding accuracy (root mean squared error of $0.80\ \text{kcal/mol}$). However, recent benchmarks have shown that current transferable MLFFs often exhibit slow convergence and limited transferability when applied to condensed-phase systems\cite{picha2025transferable,unke2024biomolecular}. Consequently, classical FFs remain the practical and reliable choice for simulations of liquids and other condensed environments.

The problem of choosing a set of partial charges, typically atom-centered, arises because they are not quantum mechanical (QM) observables; hence, there is no unique way to assign them, and the method to derive partial charges is usually decided depending on the specific application. In the context of FE calculations, a common choice is to use electrostatic potential (ESP) or restrained electrostatic potential (RESP) charges\cite{bayly1993well,woods2000restrained}. These are obtained by fitting the electrostatic potential of the molecule, usually obtained via Hartree-Fock (HF) calculations, within a shell around its van der Walls (vdW) surface. The RESP fitting considers additional restraints aimed at avoiding overpolarization and reducing conformational dependence. Since these charges rely on QM calculations, they are more accurate compared to empirical descriptors but computationally expensive; for this reason, semi-empirical methods are generally more popular, especially when needed for high-throughput workflows. The most widely used semi-empirical method is AM1-BCC\cite{jakalian2000fast,jakalian2002fast}, which performs an initial AM1 population analysis\cite{dewar1985development} and subsequently applies Bond Charge Corrections (BCC) to reproduce RESP charges. This approach inevitably renounces some of the accuracy of first-principle calculations but speeds up the charge assignment, making it possible to parametrize a large number of molecules in a reasonable amount of time (seconds to minutes per molecule, on a single CPU).

To overcome the computational cost of QM calculations, graph and convolutional neural networks have been used to predict high quality charges\cite{su2025lumicharge,wang2021deepatomiccharge} and AM1-BCC charges\cite{wang2024espalomacharge}. For the same scope, simpler architectures employing XGBoost\cite{chen2016xgboost} regression and Atom-Path-Descriptors have also been tested\cite{wang2020fast}. Although all of these studies achieved good predictions on the respective evaluation sets, the effects of their predictions on downstream calculations have not been studied. More oriented to applications to MD simulations was the work of Plé \textit{et al.}\cite{ple2023force}, where the Allegro architecture\cite{musaelian2023learning} was used to generate embeddings that were in turn used as input to multi-layer perceptrons (MLP) to predict Coulomb and dispersion interaction terms to insert into a classical FF. Here, the Allegro module was trained to get the descriptors and the output of the MLP was a set of pair-wise Coulomb contributions, which were used to construct atomic charges. Although charges constructed in this way are employable for their proposed force field, they are not directly comparable with ESP (or RESP) charges, which are more suitable to determine forces in classical MD simulations.
Another challenge in using QM derived charges in fixed charge classical FFs is their dependence on the input conformation. To address this issue, conformational ensemble approaches have been proposed\cite{basma2001solvated,reynolds1992atomic}. These approaches rely on Boltzmann averages, thus reducing the charge fluctuation within conformations; however, in doing so, they miss the charge enrichment in mutually polarized environments (e.g. water-solute systems) unless conformations are explicitly derived in those environments.

In this work, we tackle the trade-off between accuracy of partial charges and speed of their assignment, and analyze the effect that it has on absolute hydration free energy (AHFE) calculations. To do this, we first use a subset of the FreeSolv dataset\cite{mobley2014freesolv} to prove that calculations initialized with ESP charges are more accurate than those initialized with AM1-BCC charges, both in terms of root mean squared error (RMSE) and ranking (Kendall's $\tau$ and Spearman's $\rho$). The analysis of computed energies and functional groups in the respective molecules draws a non-trivial connection between the quality of the charge assignment and the quality of downstream free energy calculations. We then proceed to propose an ML method for a fast and accurate assignment of ESP charges at a high-fidelity DFT level, employing XGBoost in combination with the pretrained MACE-OFF23(L)\cite{kovacs2025mace} force field. This model requires minimal training and in a fraction of a second predicts charges that lead to energies that reproduce the ones obtained with DFT-ESP charges. Finally, we exploit the speed and accuracy of the trained model to propose the Boltzmann Percentile (BP) method to assign ESP-derived charges that are representative of the molecular conformational ensemble in the gas-phase, but favor larger charges in order to approximate polarization effects. On a final set of 30 molecules extracted from the same dataset, we show how the BP charges improve AHFE calculations of molecules characterized by conformational flexibility and polar functional groups, with respect to charges predicted on one conformation (in the following referred to as "1-shot charges"), and how both methods perform better than AM1-BCC charges.

\section{Methods}
\subsection{Computation of Partial Charges}

In this work, three methods for computing partial charges are used. The baseline is given by Electrostatic Potential (ESP) charges, which are obtained by fitting partial charges from a classical Coulomb potential to the quantum-mechanical (QM) molecular electrostatic potential (MESP). This fit is performed using the Merz-Kollman (MK) scheme\cite{singh1984approach,besler1990atomic}, which performs least squares optimization to minimize the difference between these two potentials on a set of grid points in van der Waals (vdW) shells around the molecule. This approach can generate sensibly different charges for different input conformations, and is prone to overpolarizing atoms. To avoid these problems, restrained ESP (RESP)\cite{bayly1993well,woods2000restrained} charges are often used, which perform the fitting procedure with hyperbolic restraints. Both ESP and RESP charges require the computation of the MESP, and this involves expensive QM calculations, often performed with density functional theory (DFT) methods. Semi-empirical methods aim to assign partial charges with a reduced computational cost compared to DFT. We focus on AM1-BCC charges\cite{jakalian2000fast,jakalian2002fast}, which are the most commonly used for free energy calculations. This approach combines AM1 population analysis\cite{dewar1985development} with Bond Charge Corrections (BCCs), which are aimed at reproducing RESP charges at the Hartree-Fock (HF) level of theory. These charges usually provide a good balance between speed and accuracy and, for this reason, are the common choice when simulating large systems or in high-throughput screening, when the charge assignment has to be performed on a large number of molecules.

ESP charges have been generated using Gaussian16\cite{g16} on the self-consistent density (SCF) calculated at the PBE0 level of theory\cite{adamo1999toward,ernzerhof1999assessment}, with def2-TZVP\cite{weigend2005balanced,weigend2006accurate} basis set and the D3 version of Grimme’s dispersion with Becke-Johnson damping (GD3BJ)\cite{grimme2011effect}. RESP charges were calculated by first computing the SCF and ESP charges at the HF/6-31G* level of theory \cite{petersson1988complete,petersson1991complete}, and then extracted using the Antechamber function of the AmberTools module\cite{case2023ambertools}. Finally, AM1-BCC charges were assigned through the AmberTools Toolkit Wrapper of OpenFF\cite{mobley2018escaping,jeff_wagner_2024_13375686}.

\subsection{Prediction of Partial Charges}

\begin{figure}[h!]
 \centering
 \includegraphics[width=1.0\linewidth]{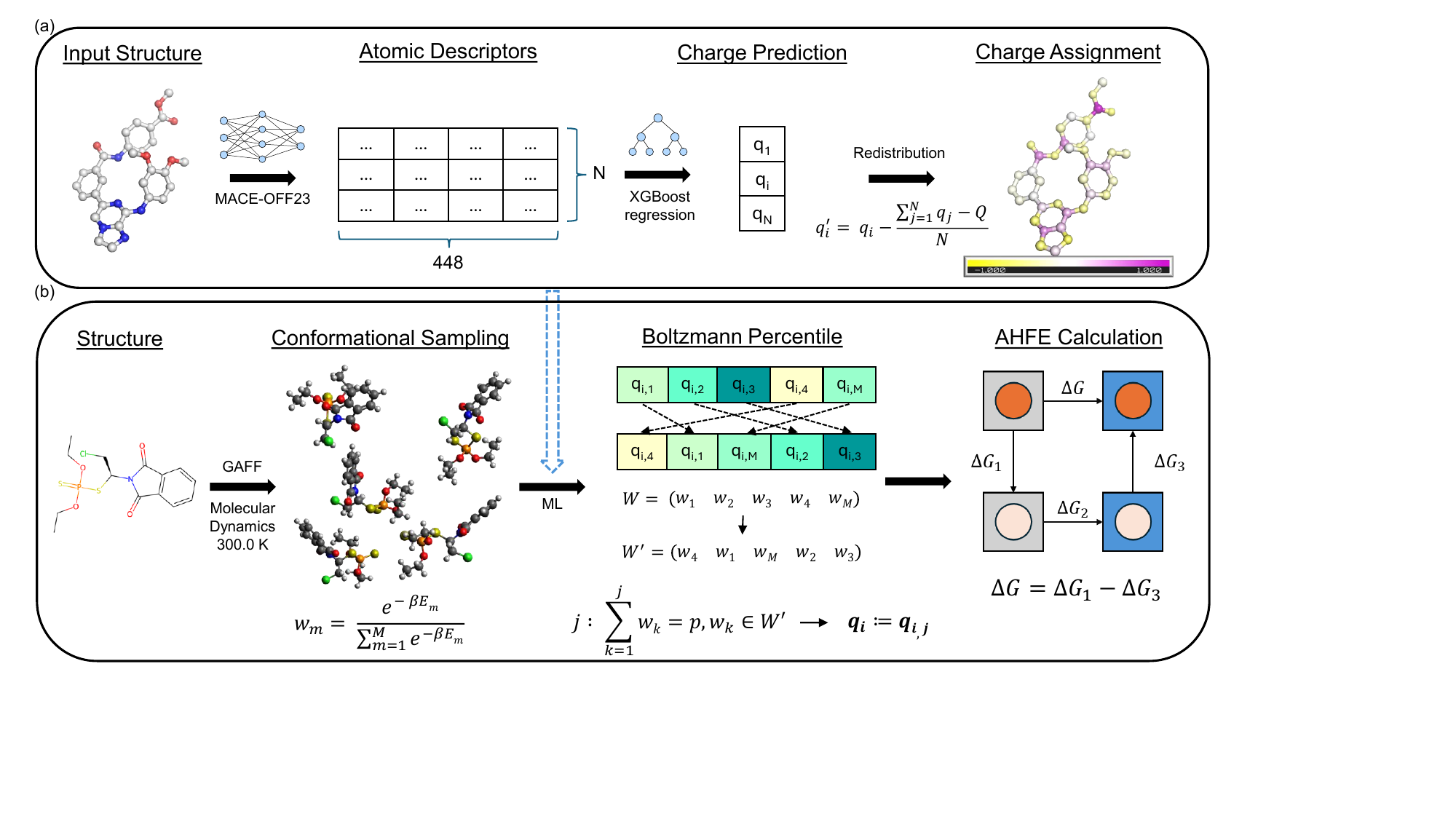}
 \caption{(a) Architecture of the model. MACE-OFF is used to generate atomic embeddings that are fed as input to an XGBoost algorithm, which ultimately predicts charges.
 (b) Boltzmann Percentile method. MD is performed on the input structure to sample conformations. Charges predicted on each conformation are then combined with Boltzmann weights to favor large charges from conformations with high probability.
 }
 \label{fig1}
 \end{figure}

Fig.~\ref{fig1}(a) shows the design of the ML model used to predict partial charges. The model takes as input atomic numbers and coordinates and uses the pre-trained MACE-OFF force field in the 23(L) version\cite{kovacs2025mace} to generate atomic descriptors. These descriptors are, in turn, used as input for an XGBoost regression algorithm\cite{chen2016xgboost} that predicts atomic charges. Charges predicted in this way are not restrained to conserve total charge, so before assignment, an additional step is performed to distribute the difference between the sum of predicted partial charges and the total charge of the molecule evenly among all atoms. 
The XGBoost model was trained on gas-phase structures from the Aquamarine (AQM) dataset\cite{medrano2024dataset}, which contains a total of 59,783 conformations from 1,653 unique drug-like molecules, with sizes ranging from 2 to 92 atoms, and containing an average of 28.2 non-hydrogen atoms belonging to elements C, N, O, F, P, S, and Cl. For each structure, a MACE descriptor was generated with the ASE package\cite{ase-paper}, and ESP charges were computed as described above. 70\% of the dataset was used for training, 20\% for validation, and 10\% for testing. Optimization was performed with Optuna\cite{akiba2019optuna}.

\subsection{Boltzmann Percentile}
To estimate representative values of atomic ESP charges from an ensemble of conformations, we introduce a \textit{Boltzmann-weighted percentile approach} (BP), as depicted in Fig.~\ref{fig1}(b).
Consider a molecule with $N$ atoms and an ensemble of $M$ conformations. Each conformation $m$ is characterized by its energy $E_m$ and the vector of atomic charges 
$$\textbf{Q}_m =(q_{1,m},q_{2,m},...,q_{N,m}).$$
Each conformation is assigned a normalized Boltzmann weight
$$w_m = \frac{e^{-\beta E_m}}{\sum_{m=1}^{M} e^{-\beta E_m}},$$
where $\beta = 1/(k_BT)$.
For each atom $i$, the set of charge values $\{q_{i,1},q_{i,2},...,q_{i,M}\}$ defines a Boltzmann-weighted distribution.
These values are first sorted in ascending order (in magnitude), and the corresponding cumulative distribution is computed as
$$C_i(q_{i,j}) = \sum_{k:q_{i,k}\le q_{i,j}} w_k$$
The \textit{Boltzmann-weighted p-th percentile} charge of atom $i$, denoted $q_{i,p}$, is then defined as the smallest charge value satisfying 
$$C_i(q_{i,p})\ge p,\quad p\in[0,1].$$
For example, $p=0.90$ corresponds to the 90th percentile of the Boltzmann weighted charge distribution for atom $i$.

This procedure is applied independently to each atom, yielding a set of representative charges $\{q_{i,p}\}_{i=1}^N$ that reflect both the thermodynamic probabilities of the sampled conformations and the tail behavior of each atomic charge distribution. After charges are assigned to each atom, the total charge surplus (or deficit) is redistributed as previously discussed.
The BP estimator thus provides an alternative to the traditional Boltzmann-weighted average, emphasizing statistically significant extreme values rather than mean behavior. This choice is justified by the fact that solute and water mutually polarize each other, leading to larger charges. A simple averaging suppresses extreme values and, as a consequence, FE calculations tend to overestimate AHFEs, as shown in Fig. S1(a) of the supplementary information (SI).

To sample conformations and compute energies, we performed molecular dynamics (MD) simulations in gas-phase using the OpenMM simulation package\cite{eastman2023openmm}. The systems were parameterized using the GAFF force field\cite{wang2004development} and equilibrated under Langevin dynamics at a temperature of $300\ \text{K}$ with a friction coefficient of $1\ \text{ps}^{-1}$. 
A time step of $2.0\ \text{fs}$ was employed for the numerical integration of the equations of motion. 
Following energy minimization, a production simulation of $1\ \text{ns}$ was carried out. 
Atomic coordinates were saved every 1000 integration steps (equivalent to $2\ \text{ps}$ per frame), resulting in a total of 500 trajectory frames, on which charges were predicted with the previously presented ML model, and the 90-th percentile was used for the assignment.

\subsection{Compound Selection and Analysis}

\begin{figure}[t!]
 \centering
 \includegraphics[width=1.0\linewidth]{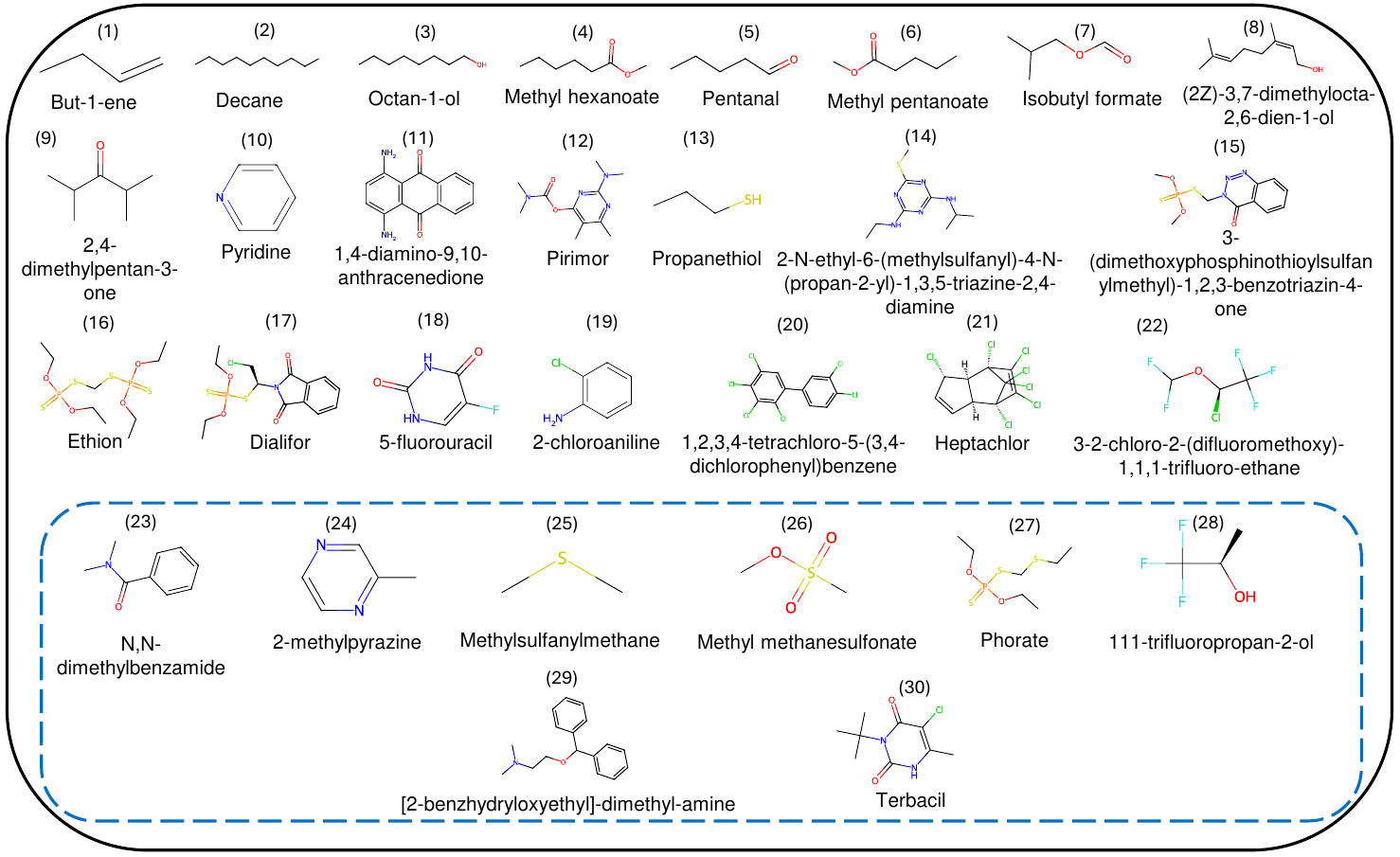}
 \caption{Systems selected for AHFE calculations. The blue box indicates the 8 extra molecules selected for the final assessment. The number in parenthesis above each molecule is an index added for easy identification in the following parity plots.}
 \label{fig2}
 \end{figure}

The compounds studied in this work were selected from the FreeSolv dataset\cite{mobley2014freesolv}, which contains experimental hydration free energies for 643 molecules. We initially selected 22 molecules with typical moieties found in drug-like molecules, to which we added 8 more molecules containing polar functional groups for the final assessment. Fig.~\ref{fig2} shows the selected molecules with names as reported in FreeSolv and 2D images drawn with RDKit\cite{greg_landrum_2023_8254217}. The selected molecules are small to medium size, with a minimum size of 11 atoms, a maximum of 41, and a mean of 23.81. The selection also spans different types of chemistry, from simple, small organic compounds to more complex aromatic heterocycles, organosulfurs, and halogenated compounds.

To connect the accuracy of the AHFE calculations with the presence of certain functional groups, we computed the total charge assigned to each group with three different assignment methods (ESP, AM1-BCC, and BP). The functional groups for each molecule are stored in the FreeSolv dataset and the specific atoms belonging to a group were found with RDKit's \texttt{GetSubstructMatches}. The total charge assigned by each method to each group, as well as the occurrence of each group, are shown in Fig~\ref{fig5}.

To compare the sensitivity of different charge assignments to the input conformation, we studied the change in atomic charges across different conformers of flexible molecules. For the sake of this analysis, we define molecules as flexible when the root mean squared deviation (RMSD) of the original structure in the input file from the geometry after geometry optimization with the UFF force field\cite{rappe1992uff} (on which charges are assigned) exceeded 2\si{\angstrom}. In the set of 22 molecules, only 6 met this criterion, of which 4 were selected as test cases for the analysis: 3-(dimethoxyphosphinothioylsulfanylmethyl)-1,2,3-benzotriazin-4-one (RMSD = 2.39 \si{\angstrom}), dialifor (2.99 \si{\angstrom}), ethion (4.00 \si{\angstrom}), and (2Z)-3,7-dimethylocta-2,6-dien-1-ol (2.51 \si{\angstrom}). For each of these molecules, we generated 9 conformers with RDKit and assigned charges with ESP, RESP, and BP as described above. We then constructed bar plots of the standard deviation across conformations of the atomic charge for each of the charging schemes (Fig~\ref{fig6}).

\subsection{Free Energy Calculations}
We have performed free energy calculations with the same exact settings but five different parameterizations of the solute's charge assignment: AM1-BCC (semi-empirical), ESP (from QM calculations), RESP (from QM calculations), 1-shot (ML-derived on the input conformation), and BP (ML-derived on the conformational ensemble).
All absolute hydration free energy (AHFE) calculations were performed using the \texttt{AbsoluteSolvationProtocol} of the OpenFE package\cite{gowers_2024_14052995}. 
This protocol performs an alchemical transformation by annihilating the solute's Coulombic interactions and subsequently decoupling its Lennard–Jones interactions from the solvent environment. 
Each transformation was sampled over 22 $\lambda$-windows, and four independent protocol repeats were executed to improve statistical reliability. 
The solute parameters were generated using the GAFF force field\cite{wang2004development}, and the systems were solvated in explicit TIP3P water\cite{jorgensen1983comparison} with a cubic box and a solvent padding of $1.5\ \text{nm}$. 
Simulations were conducted at $298.15\ \text{K}$ and 1 atm using a Langevin thermostat (collision rate $1\ \text{ps}^{-1}$) and a Monte Carlo barostat applied every 25 integration steps. 
A timestep of $4\ \text{fs}$ was employed with all bonds involving hydrogen atoms constrained, and the masses of hydrogen atoms increased. 
Non-bonded interactions were treated using the particle mesh Ewald (PME) method\cite{darden1993particle} with a real-space cutoff of $1.0\ \text{nm}$. 
For each $\lambda$-state, production simulations were run for $10\ \text{ns}$ in solvent and $2\ \text{ns}$ in vacuum, following energy minimization and equilibration phases. Free energy estimates were obtained from the multistate Bennett acceptance ratio (MBAR)\cite{shirts2008statistically} analysis as implemented in OpenFE.
All non-mentioned settings were maintained at their OpenFE defaults. 

\section{Results and Discussion}
 
\subsection{Model Validation}
\begin{figure}[t!]
 \centering
 \includegraphics[width=1.0\linewidth]{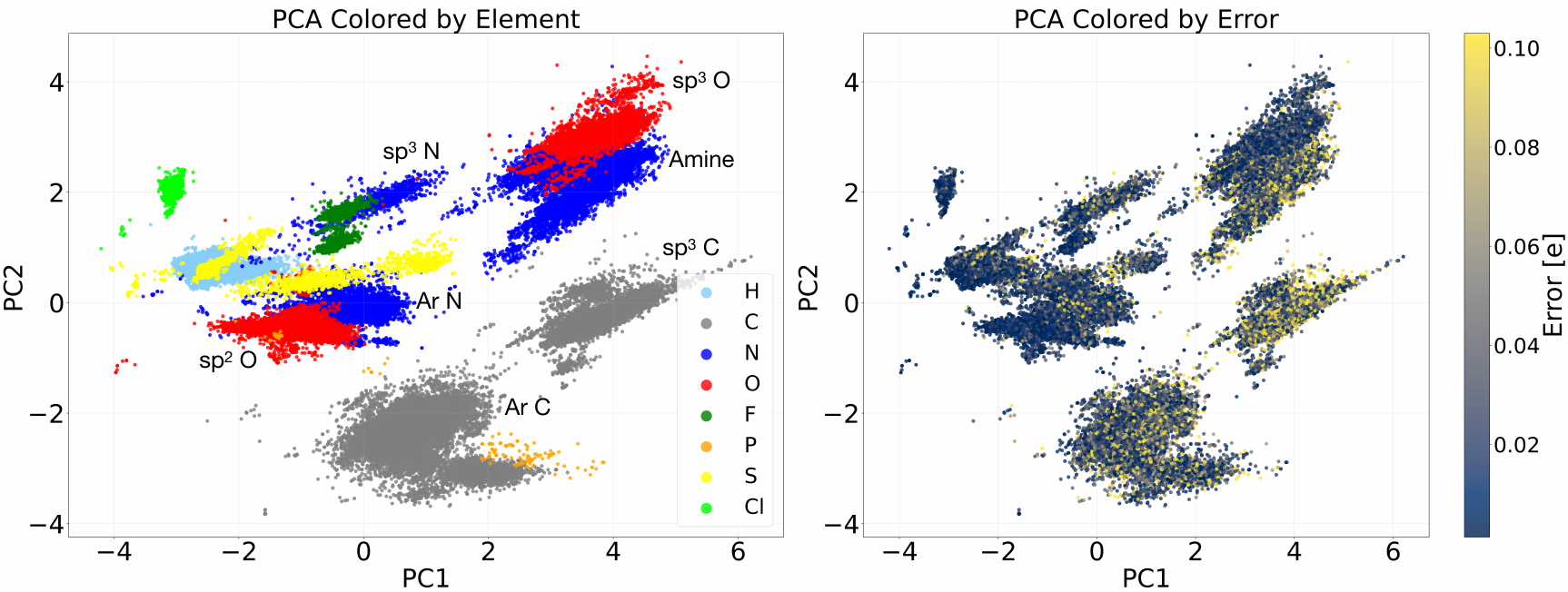}
 \caption{(left) 2D PCA projection of atomic descriptors colored by element. C, N, and O can be found in different distinct clusters, according to the atom type (as provided by OpenBabel). Carbon can be found in sp$^3$ hybridization and in Aromatic (Ar) form, while Oxygen in sp$^3$ and sp$^2$ hybridizations. N can be found in three different forms: Aromatic, Amine, and hybridized sp$^3$.
 (right) Same projection but colored by prediction error.
 }
 \label{fig3}
 \end{figure}

The root mean squared error on the test set, composed of 10\% of AQM (around 300,000 atoms) was $0.05\ \text{e}$, indicating a generally good prediction. The accuracy of the predictions is not the same for all atomic species, which is deducible from Fig.~\ref{fig3}. The left panel of the figure shows a two-dimensional principal component analysis (PCA)\cite{pedregosa2011scikit} of the atomic descriptors, colored by element. The MACE descriptors effectively group different elements in different regions of the latent space. Moreover, different atomic species (reported as OpenBabel\cite{o2011open} atom types) of the same element are also grouped separately. Interestingly, the proximity of different species is chemically reasonable. For example, amine N and sp$^3$ O are very close in the reduced space, and indeed they are both heteroatoms with lone pairs in sp$^3$ hybrid orbitals, typically bonded to carbon and hydrogen. So, they have comparable local geometries (roughly tetrahedral), and this is what is captured by the employed geometric descriptors.

The right panel of Fig.~\ref{fig3} shows the same bidimensional projection but colored by absolute prediction error. It appears clear that C and N, which have the largest number of different bonding patterns they can participate in, are the elements with larger prediction errors, while H, O, F, and Cl show very small errors. Instead, P and S are predicted with an accuracy in between the two ($0.05\ \text{e}$).
In terms of accuracy on different atomic types of the same element, charges on aromatic C and N atoms are slightly more accurate than on the same elements in hybridization $\text{sp}^3$ (Table~\ref{tab1}).
This analysis suggests that there are two main factors that determine the prediction error: a large variety of bonding patterns (C and N) and small occurrence of atoms in training data (P, S). Ulterior studies may aim at improving predictions on these elements, but as shown in the following, the current accuracy is enough to have significant improvements on AHFE calculations.

\begin{table}[t!]
    \centering
    \caption{Root mean squared error for the most common atom types in the test set, as reported by OpenBabel.}
    \resizebox{\linewidth}{!}{%
    \begin{tabular}{lcccccccccccc} 
        \toprule
        & H & Ar C & $\text{sp}^3$ C & Amine & $\text{sp}^3$ N & Ar N & $\text{sp}^3$ O & $\text{sp}^2$ O & F & P & S & Cl \\
        \midrule
        RMSE [e] & 0.02 & 0.06 & 0.09 & 0.08 & 0.07 & 0.05 & 0.03 & 0.02 & 0.01 & 0.05 & 0.05 & 0.01\\
        \bottomrule
    \end{tabular}
    }
    \label{tab1}
\end{table}

\subsection{Conformational Dependence of Molecular Charge Distribution}

\begin{figure}[t!]
 \centering
 \includegraphics[width=1.0\linewidth]{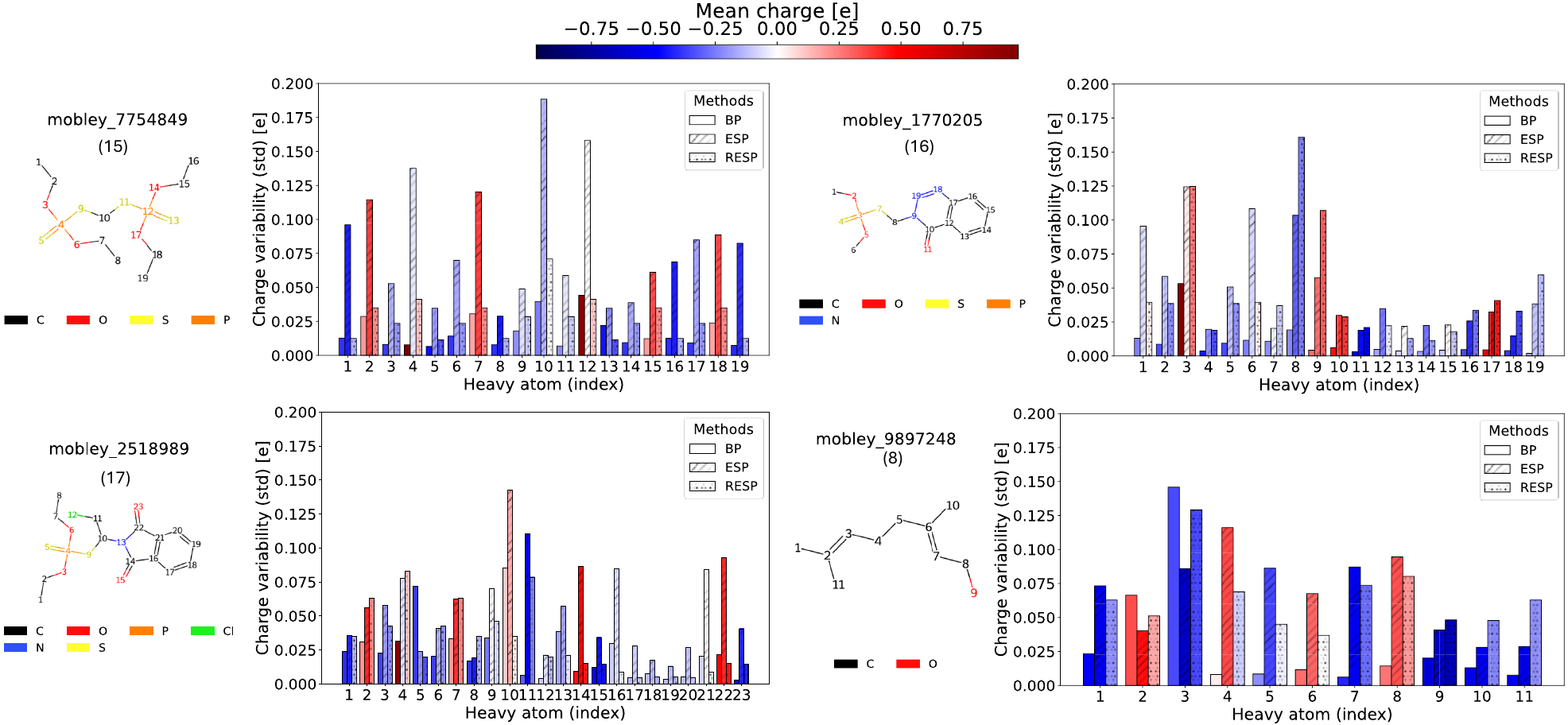}
 \caption{Variability of atomic charges for 9 different conformers for BP, ESP, and RESP method. The height of the bar indicates the standard deviation of the assigned charge, while the color indicates the mean value. }
 \label{fig6}
 \end{figure}

The sensitivity of AHFE calculations to different charge assignments found in previous works\cite{jambeck2013partial, mobley2007comparison,reynolds1992errors,osato2025evaluating} means that methods with strong conformational dependence, such as ESP, struggle in terms of reproducibility and, if the starting conformation is unrealistic, accuracy. The necessity of robustly assigning charges with different input conformations is one of the reasons RESP charges were introduced. The BP method also reduces the conformational sensitivity of ESP charges and, in the majority of cases, the assigned charges are even less variable than RESP ones.
Fig.~\ref{fig6} shows the charge variation for each compound atom across nine conformations, for four flexible compounds. As expected, the restraints in the fitting procedure ensure that RESP charges are less variable than ESP charges, \textit{i.e.} charges assigned to the same atoms of the same molecule tend to be more similar across different conformations. BP assignment not only replicates this result, but the corresponding charge distributions are even narrower than for RESP charges in most cases. This shows that for the large majority of compound atoms, the proposed method assigns charges that are less conformation-dependent than that of the well-assessed RESP method. The robustness of BP comes from using an ensemble method, which avoids structure initialization bias, when starting with an unsuitable conformation. So, rather than assigning charges to a single, possibly unrealistic, structure, the method considers several likely states.

The plots in Fig.~\ref{fig6} also suggest that the BP method overpolarizes phosphate atoms. Nevertheless, this does not seem to have major effects on AHFE calculations. The only molecules in the selection that contain P atoms are \textit{3-(dimethoxyphosphinothioylsulfanylmethyl)-1,2,3-benzotriazin-4-one}, \textit{ethion}, and \textit{dialifor}. In the next section, we discuss how BP charges lead to more accurate AHFE for these molecules compared to AM1-BCC charges.
In brief, the Boltzmann percentile method generally assigns charges that maintain the environmental awareness and the quality of ESP charges but are less dependent on the input conformation than RESP charges and thus promise to improve AHFEs and likely free energy calculations in general with respect to traditional charge assignment methods.

\subsection{Effect of Different Partial Charge Assignment Methods on AHFEs} 

\begin{figure}[h!]
 \centering
 \includegraphics[width=1.0\linewidth]{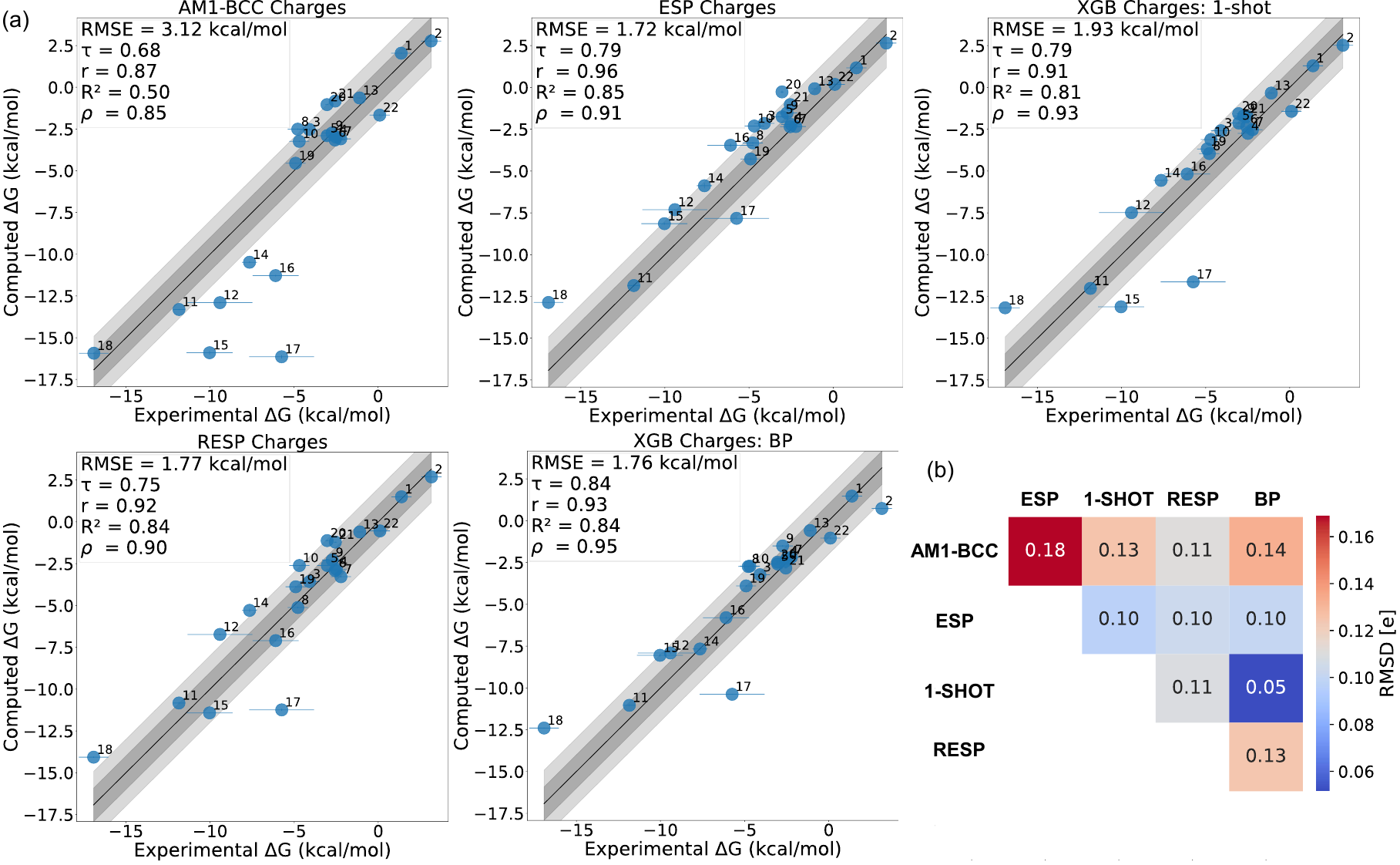}
 \caption{(a) Calculated versus experimental hydration free energies for the five charge assignment methods. Plots report root mean squared error, Kendall's $\tau$, Pearson's correlation coefficient r, $R^2$ score, and Spearman's $\rho$. Each point reports the molecular index as defined in Fig.2 for easy reference.
 (b) Pairwise root mean squared deviation of assigned charges for the 22 considered molecules.
 }
 \label{fig4}
 \end{figure}
\begin{table}[h!]
    \centering
    \caption{Root mean squared error, Kendall's $\tau$, Pearson's correlation coefficient $r$, explained variance $R^2$, and Spearman's $\rho$ of calculated values of AHFE for 22 molecules of the FreeSolv dataset with different charge assignments.}
    \begin{tabular}{lccccc} 
        \toprule
        & $RMSE\;[kcal/mol]$ & $\tau$ & $r$ &$R^2$ & $\rho$ \\
        \midrule
        AM1-BCC & 3.12 & 0.68 & 0.87 & 0.50 & 0.85 \\
        ESP     & \textbf{1.72} & 0.79 & \textbf{0.96} & \textbf{0.85} & 0.91 \\
        XGB: 1-SHOT  & 1.93 & 0.79 & 0.91 & 0.81 & 0.93 \\
        RESP    & 1.77 & 0.75 & 0.92 & 0.84 & 0.90 \\
        XGB: BP      & 1.76 & \textbf{0.84} & 0.93 & 0.84 & \textbf{0.95} \\
        \bottomrule
    \end{tabular}
    \label{tab2}
\end{table}

Fig.~\ref{fig4}(a) shows, for the 22 molecules in Fig.~\ref{fig2}, the experimental AHFEs versus those computed from AM1-BCC, ESP, 1-shot, RESP, and BP charges. The plots also report RMSE, Kendall's $\tau$, Pearson's $r$, explained variance $R^2$, and Spearman's $\rho$. These metrics are also reported in Table.~\ref{tab2}.
The highest RMSE, $3.12\ \text{kcal/mol}$, is obtained with AM1-BCC charges and coincides with the error obtained by the calculations reported in FreeSolv on the same selection of molecules. This consistency validates the calculations settings used. The lowest RMSE, $1.72\ \text{kcal/mol}$, is obtained with ESP charges and is considerably closer to the mean experimental error of the references ($0.71\ \text{kcal/mol}$). ML-predicted charges from one conformation replicate this result fairly well, only slightly increasing the error ($1.93\ \text{kcal/mol}$). The RESP and BP charges, instead, lead to an almost identical error ($1.77\ \text{kcal/mol}$ and $1.76\ \text{kcal/mol}$ respectively). In addition to the RMSE, ranking performances also improve when using ESP instead of AM1-BCC, with Kendall's $\tau$ increasing from 0.68 to 0.79 and Spearman's $\rho$ from 0.85 to 0.91. Approximately the same ranking scores are obtained with the 1-shot charge assignment, while the best are obtained with BP charges ($\tau = 0.84$, $\rho=0.95$).

\begin{figure}[t!]
 \centering
 \includegraphics[width=0.8\linewidth]{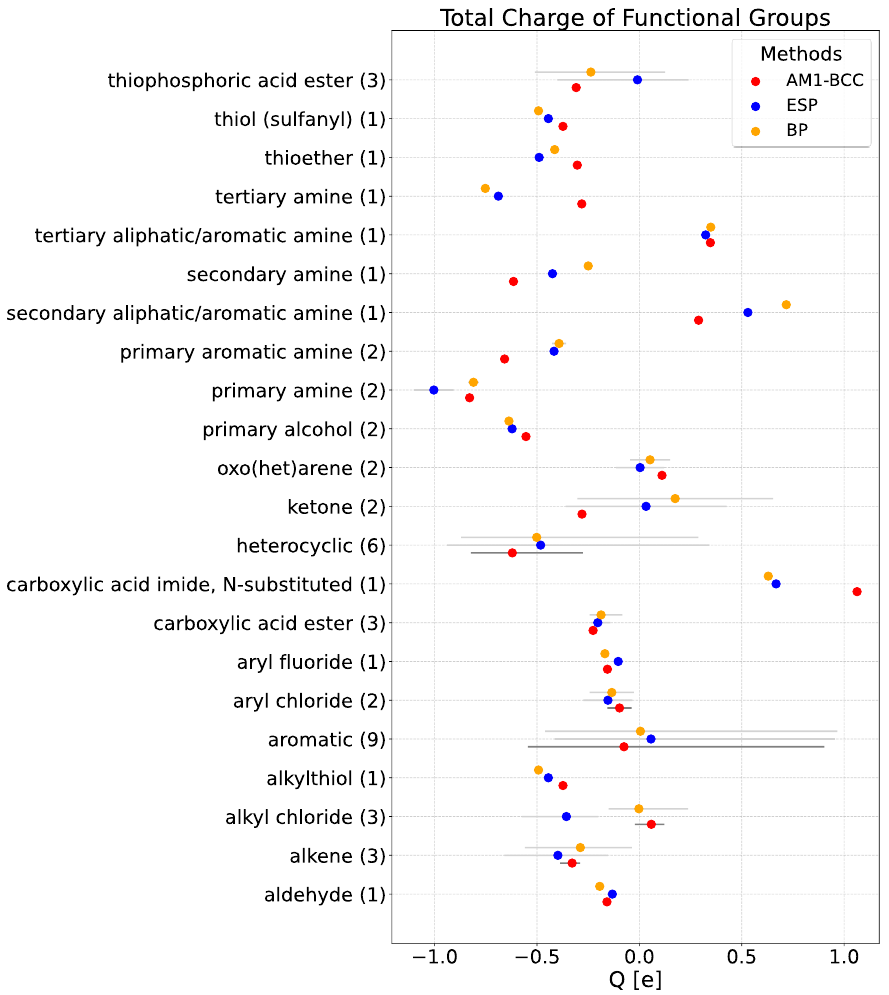}
 \caption{Total charge on functional groups of the 22 selected molecules of the FreeSolv dataset, as assigned by ESP, AM1-BCC, and the BP method. The number in parenthesis refers to the occurence of the group. The points indicate the mean value, and the bars range from the minimum to the maximum value.}
 \label{fig5}
 \end{figure}

The differences in free energies calculated from ESP and from AM1-BCC charges somehow reflect the differences between the charges themselves. In fact, as shown in Fig.~\ref{fig4}(b), the average root mean squared deviation (RMSD) between charges assigned with these two methods is large ($0.18\ \text{e}$). However, there is no clear way to predict \textit{a priori} how changes in partial charges will affect the resulting AHFE results. For example, the row corresponding to ESP in Fig.~\ref{fig4}(b) shows that 1-shot, RESP, and BP charging methods also have non-negligible differences ($0.10\ \text{e}$); nevertheless, RMSEs of calculated energies are similar. A reason for this could be that these methods, while globally assigning charges rather different from ESP, assign similar charges on the atoms that actively lead the dynamics, \textit{i.e.} on some important functional groups. 
Fig.~\ref{fig5} reports a min/max plot of the total charge assigned to the various functional groups of the selected molecules by ESP, AM1-BCC, and BP. The name of the group is followed by a number in parentheses indicating the number of occurrences of that group in the selection. For many groups, the mean values obtained with the three methods are similar, but for some AM1-BCC assign a total charge that is radically different from the other methods. It is possible to connect these groups with the outliers of the AHFE calculations reported in Fig.~\ref{fig4}(a) (defined as having an absolute error larger than $2.00\ \text{kcal/mol}$).

The poor results of calculations with AM1-BCC charges are mainly imputable to 5 outliers, corresponding to indices 14, 16, 12, 15, 17. These correspond in order to \textit{2-N-ethyl-6-(methylsulfanyl)-4-N-(propan-2-yl)-1,3,5-triazine-2,4-diamine}, \textit{ethion}, \textit{pirimor}, \textit{3-(dimethoxyphosphinothioylsulfanylmethyl)-1,2,3-benzotriazin-4-one}, and \textit{dialifor} (Fig.~\ref{fig2}).
\textit{2-N-ethyl-6-(methylsulfanyl)-4-N-(propan-2-yl)-1,3,5-triazine-2,4-diamine} is the only molecule reporting secondary amine and secondary aromatic amine groups. For both these groups, the total charge assigned to the group by AM1-BCC is notably different from the charges assigned by ESP and BP. The absolute error obtained using semi-empirical charges is $3.69\ \text{kcal/mol}$, and reduces to $0.77\ \text{kcal/mol}$ with ESP charges ($0.02\ \text{kcal/mol}$ with BP).

\textit{Ethion} contains thiophosphoric acid ester. This group also belongs to two other outliers: \textit{3-(dimethoxyphosphinothioylsulfanylmethyl)-1,2,3-benzotriazin-4-one} and \textit{dialifor}. Like the secondary amine groups, for this group, the AM1-BCC charges differ largely from the ESP and BP charges. Moreover, these two methods show a variability in charge due to different environments that the AM1-BCC method neglects. This unawareness of the local environment is directly reflected in AHFE calculations: \textit{ethion} misestimates the experimental value by $5.31\ \text{kcal/mol}$ when calculated with semi-empirical charges and by $2.82\ \text{kcal/mol}$ with ESP charges ($0.30\ \text{kcal/mol}$ with BP). The other two molecules containing this group also get lower errors when using ESP (BP) charges: \textit{3-(dimethoxyphosphinothioylsulfanylmethyl)-1,2,3-benzotriazin-4-one} from $5.80\ \text{kcal/mol}$ to $1.45\ \text{kcal/mol}$ ($1.98\ \text{kcal/mol}$ with BP), \textit{dialifor} from $9.98\ \text{kcal/mol}$ to $2.56\ \text{kcal/mol}$ ($4.64\ \text{kcal/mol}$ with BP). All of these molecules contain hypervalent phosphorus/sulfur, on which the BCC corrections were not trained extensively and have already been shown to perform poorly\cite{mobley2009predictions}.

\textit{Dialifor} contains other two functional groups for which AM1-BCC and ESP charges differ significantly: carboxylic acid imide, N-subsistuted, and alkyl chloride. The first is present only in this molecule, while the second is also present in \textit{3-2-chloro-2-(difluoromethoxy)-1,1,1-trifluoro-ethane} and \textit{heptachlor}. These two molecules are not outliers for calculations with AM1-BCC charges, but it is interesting to note that calculations with ESP charges still show lower errors: from $1.75\ \text{kcal/mol}$ to $0.17\ \text{kcal/mol}$ for the former and from $1.73\ \text{kcal/mol}$ to $1.49\ \text{kcal/mol}$ for the latter. For this group BP charges show values more similar to AM1-BCC than ESP, nevertheless they still lead to lower errors: $1.14\ \text{kcal/mol}$ for \textit{3-2-chloro-2-(difluoromethoxy)-1,1,1-trifluoro-ethane} and $0.28\ \text{kcal/mol}$ for \textit{heptachlor}.

Finally, \textit{pirimor} contains tertiary amine, on which a previous study found calculations with AM1-BCC charges more accurate than with RESP charges\cite{muddana2014sampl4}. For this molecule, the error actually reduces from $3.49\ \text{kcal/mol}$ to $1.91\ \text{kcal/mol}$ ($1.5\ \text{kcal/mol}$) when using ESP (BP) charges instead of AM1-BCC. However, a larger number of compounds with tertiary amines would be needed for a more thorough comparison in this case.

The previous analysis showed that, as expected, ESP charges can better capture the chemical environment of molecules than semi-empirical charges, and when the charges assigned with the two methods are significantly different, calculations with QM charges are always more accurate. Interestingly, these cases correspond exactly to the cases in which AM1-BCC is known to struggle: heavily halogenated alkanes, polar compounds, and hypervalent phosphorus/sulfur. The comparison with BP charges showed that even though the difference in charges may be large on the whole molecule, the Boltzmann Percentile method assigns charges that are compatible with DFT-derived charges on the relevant functional groups, and this results in similar AHFEs. This demonstrates the importance of an accurate description of the molecular charge distribution for deriving reliable AHFEs, especially in the presence of functional groups that are polar, form strong hydrogen bonds, or contain moieties that were underrepresented in the fitting of BCC corrections. Our ML model provides a sufficiently accurate electrostatic description in these cases in a fraction of a second, which is negligible with respect to DFT calculations and even shorter than AM1-BCC assignment (Table S1 in SI).

\subsection{AHFE on Extended Set of Test Molecules}
\begin{figure}[t!]
 \centering
 \includegraphics[width=1.0\linewidth]{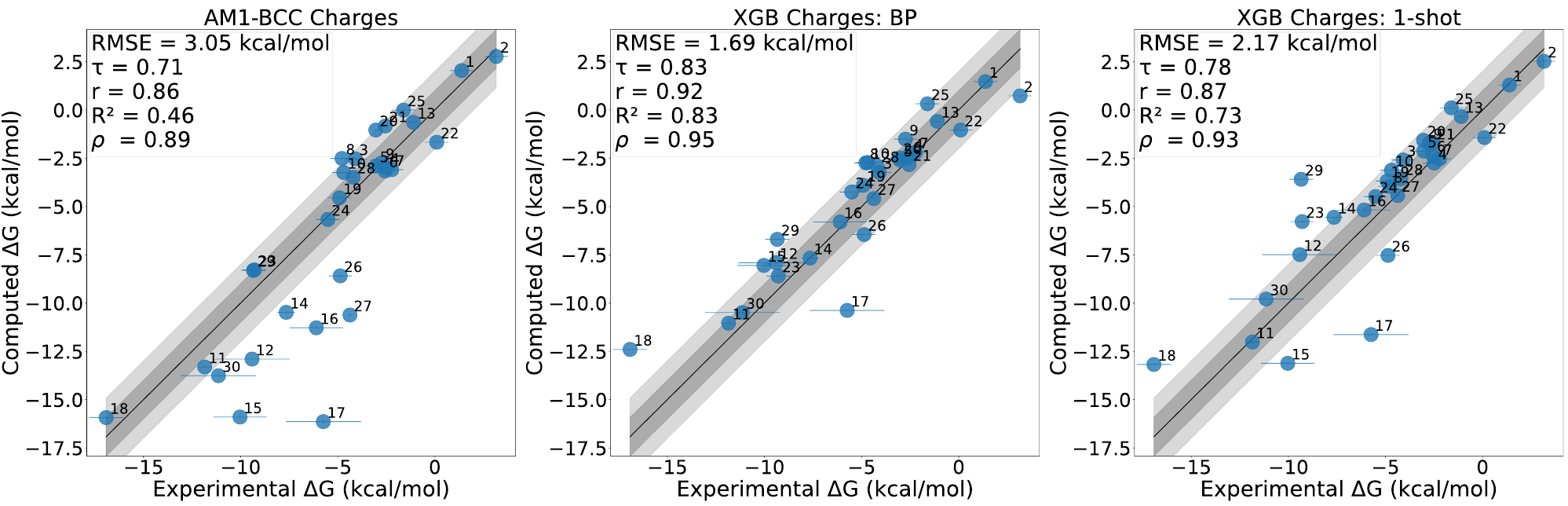}
 \caption{Parity plots of computed versus experimental AHFEs for the extended set of 30 molecules, with charges assigned with the AM1-BCC, 1-shot, and BP methods.}
 \label{fig7}
 \end{figure}

The previous analysis showed that the difficulty for semi-empirical methods to assign charges on certain moieties directly results in poor AHFE calculations. However, in the previous selection of molecules, these moieties were not highly represented. To analyze the robustness of different methods to an increased frequency of complex functional groups, we selected eight more molecules (dashed box in Fig.~\Ref{fig2}), and calculated AHFE from AM1-BCC, 1-shot, and BP charge assignments. Fig.~\Ref{fig7} shows the parity plots for the three methods on the final set of 30 molecules. 
The final metrics do not change sensibly, except for a slight increase in RMSE for 1-shot charges. Nevertheless, these charges still give more accurate calculations than AM1-BCC. Three of these 8 molecules are outliers for AM1-BCC charges: \textit{methyl methanesulfonate}, \textit{phorate}, and \textit{terbacil}. All of these molecules get good estimates using BP charges, which show only one small outlier: \textit{[2-benzhydryloxyethyl]-dimetyhl-amine}. The other two outliers for this method were already present in the previous batch of calculations and are \textit{dialifor} and \textit{5-fluorouracil}. Calculations with DFT charges also showed an outlier for \textit{fluorouracil}, indicating that QM calculations have difficulties with this compound, so ML cannot improve its description. \textit{[2-benzhydryloxyethyl]-dimetyhl-amine} and \textit{dialifor} have in common a double aromatic structure and a large number of rotatable bonds. For these compounds, assigning fixed charges is likely to be problematic. Moreover, better treatment of dispersion interactions is likely to help. The ranking performances also do not change and the BP charges ensure an extremely good ranking ($\tau = 0.83$ and $\rho = 0.95$). All together, these results, obtained on a more diverse set of molecules, are consistent with our previous findings, further validating our proposed methodology.

\section{Conclusions}
In this work, we highlight the pivotal role of the electrostatic description of molecules when performing alchemical transformations to derive absolute hydration free energies (AHFEs). The traditional approach involves force fields with atomic partial charges assigned through semi-empirical methods, typically with the AM1-BCC method, which renounces some of the accuracy of DFT calculations in exchange for a significant speed-up in the parametrization. 
The underlying assumption of this approach is that the loss stemming from less accurate charges will not drastically affect the results. However, we demonstrated that when the molecules under investigation contain functional groups that are known to be challenging to describe with semi-empirical methods, the reliability of free energy calculations is directly compromised, despite of sometimes even relatively small charge changes. To overcome this problem, we propose and make available an ML model, trained on high-fidelity DFT data (PBE0-D3(BJ)/def2-TZVP level) of comprehensive size, to reproduce QM-based ESP charges with small error ($0.05\ \text{e}$), at a fraction of their computational cost. 
The speed of the predictive model also allows us to propose a new method, namely the Boltzmann Percentile method, to assign charges that are representative of the conformational ensemble of the input molecules, thus overcoming the problem of the conformational dependence of ESP charges and avoiding a dependence on a single, potentially non-representative molecular conformation.

The proposed method produces charges that are even less variable than RESP charges, at the minimal extra cost of a short single MD simulation in gas-phase. Even charge prediction based on a single conformation already achieves better agreement with experimentally measured hydration free energies ($\text{RMSE} = 2.17\ \text{kcal/mol}$) and improves free energy ranking in comparison to calculations involving semi-empirical charges ($\text{RMSE} = 3.05\ \text{kcal/mol}$). The BP method further enhances these improvements ($\text{RMSE} = 1.69\ \text{kcal/mol}$).
Therefore, we offer a computationally efficient and reliable way to parametrize the electrostatic terms of classical force fields, which adds only a negligible extra cost to traditional workflows and significantly improves AHFE results. As a further step, suitability of our charge assignment method for ligand binding free energies will be tested. Improved reliability in free energy calculations at a computationally negligible cost and straightforward to apply, as demonstrated in this work, will contribute to improving the quality of virtual screening efforts that are routinely used in typical drug discovery projects.

\section*{Data Availability Statement}
The data underlying this study are openly available on GitHub at \href{https://github.com/mathilfiker/ml_for_charges}{this repository}. The training, validation, and test sets are available as Zenodo \href{https://doi.org/10.5281/zenodo.17790330}{repository}

\begin{acknowledgement}

The study was partially supported by the Marie Sklodowska-Curie Innovative Training Network European Industrial Doctorate grant agreement No. 956832 “Advanced machine learning for Innovative Drug Discovery” (AIDD) 

\end{acknowledgement}

\begin{suppinfo}

Time needed for charge assignment with AM1-BCC, DFT, and XGBoost on each of the 22 molecules, AHFE calculations with Boltzmann Average and Boltzmann 75-percentile, full list of calculated AHFEs with all the charge assignments.

\end{suppinfo}

\section*{Competing interests}
The authors have no competing interests to declare that are relevant to the content of this article. 
\bibliography{achemso-demo}

\end{document}